\documentclass[a4paper,12pt]{article}
\usepackage{epsfig}

\begin{document}

\def\beq{\begin{equation}}
\def\eeq{\end{equation}}
\def\bea{\begin{eqnarray}}
\def\eea{\end{eqnarray}}

\newcommand{\dedouble}{ \stackrel{ \leftrightarrow }{ \partial } }
\newcommand{\deR}{ \stackrel{ \rightarrow }{ \partial } }
\newcommand{\deL}{ \stackrel{ \leftarrow }{ \partial } }
\renewcommand{\thefootnote}{\fnsymbol{footnote}}
\rightline{} \rightline{HIP-2003-19/TH}

\rightline{ROME1-1352/2003} 
\vspace{.3cm} 
{\Large
\begin{center}
{\bf Large extra dimension effects  in Higgs boson \\ production 
at linear colliders and Higgs factories}

\end{center}}
\vspace{.3cm}

\begin{center}
Anindya Datta$^{1}$, Emidio Gabrielli$^{1}$, and Barbara Mele$^{2}$ \\
\vspace{.3cm}
$^1$\emph{
Helsinki Institute of Physics,
     POB 64, University of Helsinki, FIN 00014, Finland}
\\
$^2$\emph{Istituto Nazionale di Fisica Nucleare, Sezione di Roma,
and Dip. di Fisica, Universit\`a La Sapienza,
P.le A. Moro 2, I-00185 Rome, Italy}
\end{center}

\vspace{.3cm}
\hrule \vskip 0.3cm
\begin{center}
\small{\bf Abstract}\\[3mm]
\end{center}

In the framework of quantum gravity propagating in large extra
dimensions, the effects of virtual Kaluza-Klein graviton and
graviscalar interference with Higgs boson production amplitudes 
are computed
at linear colliders and Higgs factories.  The interference
of the almost-continuous spectrum of the KK gravitons with the
standard model {\it resonant} amplitude is finite and predictable in
terms of the fundamental D-dimensional Plank scale $M_D$ and the
number of extra dimensions $\delta$.  We find that, for $M_D\simeq 1$
TeV and $\delta=2$, 
effects of the order of a few percent could be detected 
for heavy Higgs bosons ($m_H>500$ GeV)
in Higgs
production both via $WW$ fusion in $e^+e^-$ colliders and at $\mu^+\mu^-$
Higgs-boson factories.

\begin{minipage}[h]{14.0cm}
\end{minipage}
\vskip 0.3cm \hrule \vskip 0.5cm

\section{Introduction}
In recent years much attention has been paied to theories where 
the weakness of the gravitational coupling is explained by the 
presence of {\it large} compact extra spatial dimensions,
as shown in
\cite{ADD}.
In such theories, while standard model (SM) fields are confined in the
usual 4-dimensional space, the gravity can propagate in the full high
dimensional space, and its intensity is diluted in the large volume of
the extra dimensions.

The Newton's constant $G_N$ in the 3+1 dimensional space is then
related to the corresponding Planck scale $M_D$ in the $D=4+\delta$
dimensional space by
\beq
G_N^{-1}= 8 \pi R^{\delta} M_D^{2+\delta}
\label{newton}
\eeq
where $R$ is the radius of a compact manifold assumed to be on a
torus.  According to the present limits on Newton's law
\cite{Gravity_test}, one could have $M_D\sim 1$ TeV if the number of
extra dimensions is $\delta \ge 2$.

A crucial consequence of this framework is that quantum gravity
effects could become strong at the TeV scale and measurable at future
high-energy colliders.  After integrating out the compact extra
dimensions, the effective Einstein theory in 3+1 dimensions reliably
describes the interactions of the extra-dimensional gravitons with
gauge and matter fields \cite{GRW,GRWd,GRWt}.  An essentially
continuum spectrum of massive Kaluza-Klein (KK) excitations of the
standard graviton field arises, for $\delta$ not larger than about 6.
When summing over the allowed spectrum of KK states either in the
inclusive production or in the exchange of virtual KK gravitons, the
small coupling $(E/M_P)^2$ associated to a single graviton
production/exchange (where $E$ is the typical energy of the process
and $M_P$ is the Plank mass) is replaced by the quantity
$(E/M_D)^{2+\delta}$.  Then, for $M_D\sim 1$ TeV, processes involving
gravitons could well be detected at present and future high-energy
colliders.  This possibility has been quite thoroughly explored in a
number of papers \cite{GRW}-\cite{noi}.

Regarding processes with virtual KK graviton exchange, it is well
known that in general the corresponding amplitude is divergent and not
computable in the effective theory \cite{GRW}.  In particular, the
{\it real} part of the amplitude, $Re[{\cal A}]$, needs an ultraviolet
cut-off. This means that in general the theory is not predictive in
the sector of virtual KK graviton exchange.  On the other hand, the
{\it imaginary} part of the amplitude, $Im[{\cal A}]$, is finite and
cutoff independent, being connected to the branch-cut singularity of
real graviton emission \cite{GRW}.  In a recent paper \cite{noi}, we
stressed that this can have important consequences, when considering
standard model (SM) {\it resonant} processes interfering with virtual
KK graviton exchange graphs. In fact, the corresponding interference,
that is dominated by $Im[{\cal A}]$, turns out to be {\it finite}, and
predictable in terms of the fundamental Plank scale $M_D$ and the
number of extra dimensions $\delta$.

In \cite{noi}, we applied this observation to LEP physics, and
computed the effects on the $e^+e^-\to f \bar f$ physical observables
of the interference of the virtual KK graviton-exchange amplitude with
the resonant SM amplitude at the $Z$ boson pole.  We found that,
although the corresponding impact on total cross-sections vanishes,
there are finite modifications of different asymmetries, whose
relative effect amounts, in the most favorable cases, to about
$10^{-4}$.

In the present paper, we want to extend this approach to the case of a
heavy Higgs boson ($H$) production at future linear $e^+e^-$ colliders
and $\mu\mu$ colliders. By {\it heavy}, we will imply $m_H\geq 200$
GeV.  Graviton interference effects on cross-sections turns out to be
proportional to the ratio of the total width over the mass of the
resonance state, due to the imaginary part of the SM amplitude
\cite{noi}.  Then, one expects to find remarkably more conspicuous
graviton interference effects in a heavy Higgs boson production than
in the $Z^0$ boson pole physics, due to the rapidly growing Higgs
width with the Higgs mass.  One can compare , e.g.,
$\Gamma_Z/M_Z\simeq 0.027$ with $\Gamma_H/m_H\simeq 0.072, 0.29$ for
$m_H=400, 700$ GeV, respectively.  Moreover, the imaginary part of the
graviton amplitude grows quite rapidly with the process c.o.m. energy.

\begin{figure}[h]
\centerline{\hspace*{3em}
\epsfxsize=10cm\epsfysize=4.0cm
                     \epsfbox{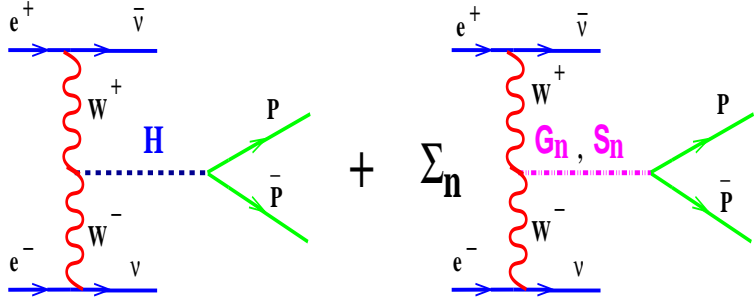}
}
\caption[]{\em Feynman diagrams of the processes in 
Eq.~(\ref{eeWWH}).}
\label{feynman_diag}
\end{figure}
At linear $e^+e^-$ colliders \cite{Aguilar-Saavedra:2001rg}, we
consider the Higgs production via vector boson fusion with the
subsequent $H$ decay into pairs of heavy particles
\beq
e^+ e^- (WW) \to \nu \bar{\nu} H  \to \nu \bar{\nu} WW, \; \nu \bar{\nu}ZZ, \;
\nu \bar{\nu}t \bar t \, .
\label{eeWWH}
\eeq
Feynman diagrams for these processes are presented in
Figure~\ref{feynman_diag}. This process is one of the dominant $H$
production mechanism at linear colliders, and becomes the main one at
$\sqrt{S} \geq$ 500 GeV \cite{higgs}. The same initial and final states
can be mediated by a continuous spectrum of KK spin-2 gravitons,
$G_n$, and spin-0 graviscalars, $S_n$, in the $s$ channel \footnote{We
neglect here possible $t$ channel amplitudes, that give subdominant
contributions in the following.}

\beq
\sum_n \left\{ e^+ e^- (WW) \to \nu \bar{\nu}\, G_n^\ast, \; \nu \bar{\nu}\, 
S_n ^\ast  
\to \nu \bar{\nu} WW, \; \nu \bar{\nu} ZZ, \; \nu \bar{\nu} t \bar t \right\}
\, .
\label{eeWWG}
\eeq

These processes at energies $\sqrt S \gg M_W$ can be reliably
treated in the effective-$W$  approximations \cite{dawson},
by convoluting the cross-sections for the subprocesses
\beq
WW \to H \to WW, ZZ, t \bar t \, , 
\label{prou}
\eeq
and  
\beq
 \sum_n  \left\{WW \to G_n ^\ast, \; S_n ^\ast \to WW, ZZ, t \bar t\right\}
\label{prod}
\eeq 
with the appropriate $W$ distributions in the electron beam
(same for $ZZ$ initiated processes).

On the Higgs boson resonance, the interference of the processes
eqs.(\ref{prou}) and (\ref{prod}) will be dominated by the
imaginary part of the graviton/graviscalar amplitude, that, as
discussed above, is finite and predictable in terms of $M_D$ and
$\delta$.

The aim of the present paper is to determine the amount by which the
Higgs production cross-sections and distributions can be affected by
the interference with the KK graviton/graviscalar amplitude.  We then
discuss the possibility to measure such an effect (and, hence, to find
a footprint of a large extra dimension theory) at realistic linear
collider machines.

In the last part of the paper we also analyze KK graviton/graviscalar
interference effects in Higgs production at a possible
$\mu \mu$ collider acting as a Higgs boson factory \cite{hfactory},
through the channels
\beq
\mu^+ \mu^- \to H \to WW, ZZ, t \bar t \; ,
\eeq
and
\beq
\sum_n  \left\{\mu^+ \mu^- \to G_n ^\ast, \; S_n^\ast \to WW, ZZ, t \bar t \right\} \; .
\eeq
This process would presumably be affected by smaller theoretical
uncertainty, and could provide an even more sensitive probe to large
extra dimension effects.

\section{The virtual-graviton exchange amplitude}

We are interested in computing the interference of the exchange of a
virtual KK spin-2 graviton and a virtual KK spin-0 graviscalar with a
resonant SM scattering amplitude. Hence, we will analyze in particular
an $s$-channel KK exchange amplitude.  We will follow in this section
the approach of \cite{GRW}.

The graviton-mediated
scattering amplitude in the momentum space 
is obtained by summing over all KK modes 
\beq
{\cal A} =\frac{1}{\bar{M}_P^2}\sum_n \left\{ T_{\mu\nu}
\frac{P^{\mu\nu\alpha\beta}}{s-m_n^2}T_{\alpha\beta}
+\frac{1}{3}\left(
\frac{\delta-1}{\delta+2}\right)\frac{T^{\mu}_{\mu}
T^{\nu}_{\nu}}{s-m_n^2}\right\} \, ,
\label{amplitude}
\eeq
where $\bar{M}_P$ is the reduced Plank mass ($\bar{M}_P=M_P/\sqrt {8
\pi}$), and $T_{\mu\nu}$ is the energy-momentum tensor of the
scattering fields.  The first and second terms in Eq.~(\ref{amplitude})
corresponds to graviton and graviscalar exchanges respectively (here
$m_n$ represents both the graviton and graviscalar masses without loss
of any generality).  In the unitary gauge, the projector of the
graviton propagator, $P^{\mu\nu\alpha\beta}$, is given by
\beq
P^{\mu\nu\alpha\beta}=\frac{1}{2}\left(\eta^{\mu\alpha}\eta^{\nu\beta}
+\eta^{\mu\beta}\eta^{\nu\alpha}\right)-
\frac{1}{3}\eta^{\mu\nu}\eta^{\alpha\beta}
+\dots
\eeq
where $\eta^{\mu\nu}$ is the Minkowski metric. Dots represent terms
proportional to the graviton momentum $q_{\mu}$, that, being
$q^{\mu}T_{\mu\nu}=0$, give a vanishing contribution to the amplitude.
The trace of $T_{\mu\nu}$ is nonvanishing only for massive initial and
final states.

Since the energy-momentum tensors do not depend on 
KK indices, one can perform the sum (over $n$) irrespective  of the
scattering process, and Eq.~(\ref{amplitude}) becomes 
\beq
{\cal A}={\cal S}(s) 
{\cal T},~~~~
{\cal S}(s)=\frac{1}{\bar{M}_P^2} 
\sum_n \frac{1}{s-m_n^2},~~~~
{\cal T}=T_{\mu\nu}T^{\mu\nu}- \frac{1}{\delta+2} T^{\mu}_{\mu}
T^{\nu}_{\nu}
\label{reducedA}
\eeq
In the continuum approximation for the
KK graviton spectrum, one then obtains  
\beq
{\cal S}(s)=\frac{1}{M_D^{2+\delta}} \int d^{\delta} q_T 
\frac{1}{s-q_T^2} = \frac{\pi^{\frac{\delta}{2}}}{M_D^4}
\Gamma(1-\frac{\delta}{2})
\left(-\frac{s}{M_D^2}\right)^{\frac{\delta}{2}-1}
\eeq
where we  assumed $m_n^2=q_T^2$, with $q_T$ the graviton momentum 
orthogonal to the brane.

In the interference with a resonant  
amplitude, only  $ Im[{\cal S}(s)]$ will contribute,
with 
\beq
Im [{\cal S}(s)]=-\frac{\pi}{M_D^{2+\delta}}\frac{S_{\delta-1}}{2}
s^{\frac{\delta-2}{2}}
\label{ImS}
\eeq
where $S_{\delta-1}$ is the area of the $\delta$ sphere. For $\delta=2n$,
$S_{\delta-1}=2\pi^n/(n-1)!$ and,  for $\delta=2n+1$,
$S_{\delta-1}=2\pi^n/\prod^{n-1}_{k=0} (k+\frac{1}{2})$, 
with $n$ integer.  Hence, imaginary part of the amplitude is finite and
predictable, only depending on the D-dimensional Plank scale $M_D$
and on the number of extra dimensions $\delta$. It also grows quite
rapidly with $\sqrt{s}$ for $\delta > 2$.

On the other hand, as discussed in the previous section, divergences
in general arise in the real part of ${\cal S}$. They can be
regularized by introducing an external cut off that spoils the
predictivity of the theory, parametrizing unknown new-physics
contributions in the ultraviolet region.  This feature makes virtual
graviton exchange processes phenomenologically less interesting than
the real graviton production processes, unless the imaginary part of
the amplitude can be separated. This is exactly what happens when
considering the graviton interferences with resonant SM scattering
amplitudes.  The latter case is indeed the subject of our previous
work in \cite{noi} and of the present paper.

\section{Interference effects in the $WW$ partonic cross-sections}

In this section, we determine the effects on the angular distributions
and cross-sections of the $WW$ fusion Higgs production processes in
Eq.~(\ref{eeWWH}) arising from their interferences with the
corresponding KK graviton/graviscalar exchange processes in
Eq.~(\ref{eeWWG}).  We start from the {\it partonic} $WW$ initiated
amplitude for the processes
\beq
WW \to H \to WW, ZZ, t \bar t \, , 
\label{WWH}
\eeq
and  
\beq
 \sum_n  \left\{WW \to G_n ^\ast, \; S_n^\ast \to WW, ZZ, t \bar t\right\} \, ,
\label{WWG}
\eeq 
we compute their interferences, and then convolute them (along with
the corresponding SM cross sections as in Eq.~(\ref{WWH}) with the 
effective-$W$ distributions in the initial electron/positron beams.

We notice that due to the different spin properties of the Higgs
($s=0$), graviton ($s=2$) and graviscalar ($s=0$) 
intermediate states, only the graviton  
will have a nontrivial impact on the angular distribution.
On the other hand, the latter effect will vanish in the
total cross section, since different spin amplitudes turn out
to be orthogonal.
 An analogous
effect can be observed in the $Z$ boson - graviton interference in
\cite{noi}.  

\noindent
The initial $W$ polarizations that are relevant for Higgs
production are the ones where both the $W$'s are either transverse
(with opposite polarization projection) or longitudinal.  We call the
two combinations, $\lambda=T$ and $\lambda=L$, respectively.

If $P$ stands for one of the possible
final particles $W$, $Z$ and $t$ in Eq.~(\ref{WWH}),
the polarization dependent angular distribution 
for the process $W^+W^- \rightarrow P \bar P$ via Higgs exchange
plus interference effects with the graviton/graviscalar
mediated scattering reads, near the Higgs boson pole (i.e., for 
$|\sqrt{\hat{s}}-m_H|< \Gamma_H$)
\footnote{Contributions coming from the real part of the 
amplitudes are suppressed by terms of order $|\hat{s}-m^2_H|/m^2_H$
in this case.}
is given by
\bea
{\frac{d \sigma^P_{\lambda}}{d \cos{\theta}}} &=& 
\frac{\bar{\sigma}^P_{\lambda}}{2} \;   
\left\{1 + \Delta^P_{0} + \Delta^{P}_{2,\lambda}
 \;(1 - 3\cos^2{\theta}) \right\} \, ,
\label{master_eqn}
\eea
where $\theta$ is the polar angle of one of the final particle with respect 
to the beam in the c.o.m. frame.
Nonvanishing coefficients $\Delta^P_0$ and $\Delta^{P}_{2,\lambda}$ 
arise from the interference of the Higgs exchange amplitude
with the graviscalar and graviton exchange amplitudes, respectively.
 In particular, a nonvanishing graviton contribution alters the 
 $\cos{\theta}$ independent distribution predicted by the SM Higgs exchange.
$\bar{\sigma}^P_{\lambda}$
stands for the SM Higgs-exchange total cross section
for the process $W^+W^- \rightarrow P \bar P$, 
\beq
\bar{\sigma}^P_{\lambda} = 
\frac{1}{16 \pi \hat{s}}\;\frac{g^4 m_W^4  \xi_P}{(\hat{s} - m_H^2)^2 + 
m_H^2 \Gamma _H^2} \; \sqrt{\frac{\hat{s} - 4m^2_P}{\hat{s} -4m_W^2}}\; 
\rho^P_{\lambda}\left(\frac{\hat{s}}{m_W^2}\right).
\label{SM_eqn}
\eeq
In the last equation, $g$ is the electroweak $SU(2)$ coupling,
$m_P$, $m_W$, and $m_H$ stand for the $P$, $W$, and $H$ masses,
respectively, $\Gamma _H$ is the total Higgs width,
$\sqrt{\hat{s}}$ is the total energy of the initial $W$'s in their 
c.o.m. frame, and $\xi_P$ are numerical coefficients 
($\xi_t = \xi_W = 1$, and $\xi_Z = \frac{1}{2}$). 
The functions $\rho^P_{\lambda}(x)$ are given by
\bea
\rho^t_{L}(x) &=& \frac{(x - 2)^2}{4} \rho_T^t(x)~,~~~
\rho^t_{T}(x) =\frac{3}{2} r_t ( x - 4r_t) \nonumber \\
&& \nonumber \\
\rho^{V}_{L}(x) &=& \frac{(x -2)^2}{4}\rho^{V}_T(x)~,~~~
\rho^{V}_T(x) =
\frac{(x^2 - 4xr_V + 12r_V^2)}{4} \nonumber 
\eea
where, $V=(W,Z)$, $r_t = m_t^2/m_W^2$, 
$r_W=1$, and $r_Z=m_Z^2/m_W^2$.

Finally, we present the expressions of the coefficients
$\Delta^P_0$ and $\Delta^{P}_{2,\lambda}$ in Eq.~(\ref{master_eqn}),
arising from the interference of the SM Higgs exchange amplitude 
near the Higgs pole with the
imaginary part  [cf. Eq.~(\ref{ImS})] of  the graviscalar/graviton
exchange amplitude
\bea
\Delta_{0}^P&=& R_\delta\, c_P\left(\frac{\delta -1} {\delta +2}\right)~,~~~
\Delta_{2,\lambda}^{P} = R_\delta \, f_{\lambda}^P
\left(\frac{\hat{s}}{m_W^2}\right)
 \nonumber \\ 
R_{\delta} &=&
\frac{S_{\delta -1} \pi}{4\sqrt{2} G_F M_D^2} \left( 
\frac{\sqrt{\hat{s}}}{M_D} \right)^{\delta} 
\left( \frac{m_H \Gamma_H}{\hat{s}}\right) 
\label{rrr}
\eea
where $c_P$ are numerical coefficients
($c_W = \frac{4}{3}$, $c_Z = \frac{2}{3}$, and $c_t = \frac{4}{3}$),
and
the functions $f^P_{\lambda}(x)$s are defined as
\bea
f^t_{L}(x) &=& - \frac{1}{2}\left(\frac{x+4}{x - 2}\right)f^t_T(x)~,~~~
f^t_{T}(x) = -\frac{4}{3} \nonumber \\
&& \nonumber \\
f^{V}_{L}(x) &=& -\frac{1}{2} \left(\frac{x+4}{x-2}\right)f^{V}_{T}(x)~,~~~
f^{V}_{T}(x) = \frac{2}{3}
\frac{(x - 4r_{V})(x + 6r_{V})}
{(x^2 - 4xr_{V} + 12r_{V}^2)} \, . \nonumber \\
\eea
We recall that interference effects arising from the {\it real} part of the 
amplitudes (that we are neglecting) are suppressed by terms of order
$|\hat{s}-m^2_H|/m^2_H$ on the Breit-Wigner resonance.

When convoluting the {\it partonic} $WW$ cross sections with 
the $W$'s effective
fluxes in the collider beams, 
it will be useful to approximate the Breit-Wigner propagator
in Eq.~(\ref{SM_eqn}) by a Dirac delta function
\beq
\frac{1}{(\hat {s} - m_H^2)^2 + m_H^2 \Gamma _H^2} \longrightarrow \frac{\pi}
{m_H \Gamma_H}\;\delta(\hat{s} - m_H^2) \, .
\label{dirac}
\eeq

A few basic features of the distribution in Eq.~(\ref{master_eqn})
can  be discussed even before making the convolution with the 
$W$'s fluxes.
First of all, as anticipated, the spin structure of the intermediate states
determine a flat (Higgs-like) angular distribution for the graviscalar 
interference  contribution, affecting the total cross section
by an amount $\Delta^P_0\times \sigma_{SM}$.
On the other hand, the spin-2 gravitons give rise to a 
$(1 - 3\cos^2{\theta})$ angular distribution in the $WW$ c.o.m frame,
that gives a vanishing result on the total cross section.
Nevertheless, an angular analysis of the final state will
reflect the nontrivial impact of the $(1 - 3\cos^2{\theta})$ 
distribution in the $WW$ c.o.m system
on the laboratory-frame angular characteristics.
For instance, some angular cut on the directions of the final states $P$
with respect to the electron/positron beams will originate
a non null effect (weighted by the coefficients $\Delta^{P}_{2,L}$
and $\Delta^{P}_{2,T}$) in the integrated cross sections.

In order to establish the general relevance of the present effects,
it is of course crucial to analyze the numerical values 
of the coefficients $\Delta^P_0$ and $\Delta^{P}_{2,\lambda}$
for interesting cases of the model parameters.
In Tables~\ref{table1} and \ref{table2} the most favorable 
(experimentally allowed) case of $M_D = 1$ TeV with $ \delta = 2 $
is presented for the processes $W^+W^- \rightarrow W^+W^- $ and
$W^+W^- \rightarrow t \bar t$, respectively.
The coefficients $\Delta^P_0$ and $\Delta^{P}_{2,\lambda}$
have been evaluated  at
$\sqrt{\hat{s}} = m_H$, and are presented for different Higgs masses
(with  $m_H\leq 800$ GeV) and corresponding total widths.
The values of relevant SM parameters in the computation
 are set as in \cite{PDG}.
The case $W^+W^- \rightarrow ZZ$ can be easily inferred from 
Table~\ref{table1}, since $\Delta^Z_0= \Delta^W_0 /2$ and, apart from
$o(M_Z^2/M_W^2-1)$
effects,
$\Delta^{Z}_{2,\lambda}=\Delta^{W}_{2,\lambda}$.

\begin{table}[ht]
\begin{center}
\begin{tabular}{|c|c|c|c|c|}
\hline
$m_H (GeV)$ & $\Gamma_H (GeV)$ & $\Delta_{2,T}^{W}$ & $\Delta_{2,L}^{W}$ & 
$\Delta_{0}^W$\\
\hline
200&1.4& $5.9 \times 10^{-5}$&$-7.1\times 10^{-5}$ & $2.8\times 10^{-5}
$  \\
\hline
300&8.5 & $ 6.7\times 10^{-4}$&$-5.1\times 10^{-4}$ & $2.6\times 10^{-4}
$  \\
\hline
400& 29.& $2.8\times 10^{-3}$&$-1.7\times 10^{-3}$ & $1.1\times 10^{-3}
$  \\
\hline
500&67.& $7.7\times 10^{-3}$&$-4.4\times 10^{-3}$&$3.3\times 10^{-3}$ \\
\hline
600& 123.& $1.6\times 10^{-2}$&$-9.0\times 10^{-3}$&$7.3\times 10^{-3
}$\\
\hline
700& 200.& $3.0 \times 10^{-2}$&$-1.6\times 10^{-2}$&$1.4\times 10^{-2}$\\
\hline
800&309. & $5.2 \times 10^{-2}$&$-2.7\times 10^{-2}$ & $2.4\times 10^{-2}
$  \\
\hline
\end{tabular}
\end{center}
\caption{Numerical values of  
$\Delta_{2,T}$,
$\Delta_{2,L}$ and $\Delta_0$ for various 
Higgs masses if $\delta =$ 2 and $M_D=1$ TeV, for the process 
$W^+W^- \rightarrow W^+W^-$.}
\label{table1}
\end{table}

\begin{table}[ht]
\begin{center}
\begin{tabular}{|c|c|c|c|c|}
\hline
$m_H (GeV)$ & $\Gamma_H (GeV)$ & $\Delta_{2,T}^{t}$ & $\Delta_{2,L}^{t}$ & 
$\Delta_{0}^t$\\
\hline
400& 29.& $-4.5\times 10^{-3}$&$2.7\times 10^{-3}$ & $1.1\times 10^{-3}$  \\
\hline
500&67.&$-1.3\times 10^{-2}$&$7.8\times 10^{-3}$&$3.3 \times 10^{-3}$ \\
\hline
600& 123.& $-2.7\times 10^{-2}$&$1.6\times 10^{-2}$&$7.3\times 
10^{-3}$\\
\hline
700& 200.& $-5.6\times 10^{-2}$&$3.0 \times 10^{-2}$&$1.4\times 10^{-2
}$\\
\hline
800&309.& $ -9.7\times 10^{-2}$&$5.0\times 10^{-2}$ & $2.4\times 10^{-2}
$  \\
\hline
\end{tabular}
\end{center}
\caption{Numerical values of $\Delta_{2,T}$,
$\Delta_{2,L}$ and $\Delta_0$ for various 
Higgs masses if $\delta =$ 2 and $M_D=1$ TeV, for the process $W^+ W^- 
\rightarrow t \bar t$.}
\label{table2}
\end{table}

The leading dependence on the Higgs mass of the coefficients
$\Delta^P_0$ and $\Delta^{P}_{2,\lambda}$ arises from the 
$R_{\delta}$ behavior as a function of $m_H$ and $\Gamma_H$.
In general, from Eq.~(\ref{rrr}), on the Higgs peak,
one has 
\beq
R_{\delta} \sim
\frac{S_{\delta -1}}{G_F M_D^2} \left( 
\frac{m_H}{M_D} \right)^{\delta} 
\left( \frac{\Gamma_H}{m_H} \right) \, .
\label{rrd}
\eeq
At $\delta=2$, $R_{2}\sim m_H \Gamma_H$, and this largely explains  the
increase of $\Delta^P_0$ and $\Delta^{P}_{2,\lambda}$ with $m_H$ observed
in Tables~\ref{table1} and \ref{table2}.
One can note that for  $m_H>500$ GeV most of the coefficients are quite large,
and could have an impact on the measurable cross-sections.
At $m_H=800$ GeV, all the coefficients amount to a few percent.
The most striking one seems to be the graviton-interference case
in the top quark
channel, that gives $\Delta^{t}_{2,T}\simeq -0.1$ at $m_H\simeq 800$ GeV.

On the other hand, increasing the Planck scale $M_D$ can quite affect
the coefficient values considered above.
From Eq.~(\ref{rrd}), one has 
$R_{\delta}\sim 1/M_D^{2+\delta}$.
Increasing $M_D$ by a factor 2 would imply, for instance,
 a reduction by a factor
about 1/16 on the coefficients values shown in 
Tables~\ref{table1} and \ref{table2}.

\section{Interference effects in the $e^+e^-$ cross-sections}
In the previous section, we have shown that, after integrating 
over the full range of $\cos \theta$ the $W^+W^- \rightarrow P \bar P$
angular distribution,   
the {\it graviton} interference, weighted by the function
$(1 - 3\cos^2{\theta}) \;$, vanishes. 
Only {\it graviscalar}-interference effects survive,
affecting the total cross sections by a factor $(1+\Delta^P_0) \;$.
The latter will modify the corresponding total cross-sections
in $e^+e^-$ collisions. 
In order to pin down the {\it graviton}-interference
coefficients $\Delta^{P}_{2,\lambda} \;$, one should instead optimize
the angular analysis of the process
by  defining proper strategies (like angular cuts or new asymmetries)
that can enhance the graviton contribution (cf. \cite{noi}).
To this end, it is crucial to consider the laboratory-frame
angular distribution for the complete process
$e^+ e^- (WW) \to  \nu \bar{\nu} P\bar P \;$,
that can be obtained by properly
boosting  the subprocess $W^+W^- \rightarrow P \bar P$
according to the initial $WW$ fluxes in the electron/positron beams.

In a laboratory frame where the initial $WW$
systems moves with velocity $\beta$, the $W^+W^- \rightarrow P \bar P$ 
angular distribution  in Eq.~(\ref{master_eqn}) 
becomes
\bea
{\frac{d \sigma^P_{\lambda}}{d \cos{\theta_L}}} &=& 
\frac{\bar{\sigma}^P_{\lambda}}{2} \;   
\left[ 1 + \Delta_{0}^P + \Delta^P_{2,\lambda}
 \;{\cal F} (\theta_L, \beta)\right]\;{\cal J}(\theta_L, \beta) \; ,
\label{lab_eqn}
\eea
where
\beq
{\cal F}(\theta_L, \beta) \equiv 1 - 3\left(\frac{\cos{\theta_L} 
- \beta}{1 - \cos{\theta_L}
\beta} \right)^2 ;~~~~{\cal J}(\theta_L, \beta) \equiv \frac{1 - \beta^2}
{(1 - \beta \cos{\theta_L})^2} \; ,
\eeq
and 
$\theta_L$ is the $P$ scattering angle in laboratory frame. 
Above, we  have neglected terms of order $m_W^2/E_e^2$.

We then fold the above {\it partonic} cross sections with the probabilities
$P^{W}_{\lambda}(x)$ of emitting from an $e^+ (e^-)$ beam a real $W$ with
polarization $\lambda$ and fraction of the beam momentum $x$.  The  
$e^+ e^- (WW) \to  \nu \bar{\nu} P\bar P$ differential
cross-section can be written  as
\beq
\frac{d \sigma^P_{ee}(S)}{d \cos{\theta_L}}
= \Sigma_{\lambda=L,T}\;\int dx_1 \;dx_2 \; 
\left\{
P_{\lambda}^{W} (x_1)
\;P_{\lambda}^{W}(x_2)\; \frac{
d\sigma^P_{\lambda}(\hat s)}{d\cos\theta_L}\right\}
\label{eqnee}
\eeq
where $\hat{s} = x_1 x_2 S \;$ and 
$\sqrt{S}=2E_e$, with $E_e$  the energy of the beam in the laboratory
c.o.m. system.
For the $W$-fluxes, we assume the expressions \cite{dawson}
 \bea
P_{T}^{W} (x) &=& \frac{g^2}{64 \pi^2}\;
\frac{x^2 + 2(1 - x)}{x}\log{\left(\frac{S}{m_W ^2}\right)}
\nonumber \\
P_{L}^{W} (x) &=&\frac{g^2}{32 \pi^2}\;\left(\frac{1 -x}{x}\right) \; .
\eea

We now can have the laboratory-frame
angular distribution for the complete process
$e^+ e^- (WW) \to  \nu \bar{\nu} P\bar P \;$,
by convoluting Eq.~(\ref{lab_eqn})     with the $W$-fluxes.
By introducing the variables
\beq
\tau = \frac{m_H^2}{S},~~~~~
\beta = \frac{x^2 - \tau}{x^2 + \tau}, ~~~~~
r_H = \frac{m_H^2}{m_W^2}\, ,
\eeq
and the definition
\beq
N_P = \frac{g^4 m_W^4  \xi_P}
{32 S m_H ^3 \Gamma _h} \; \sqrt{\frac{r_H - 4r_P}{r_H -4}}\; \rho^P_{T}(r_H)
\, ,
\eeq
and, by making use of Eq.~(\ref{dirac}), 
one has from Eqs.~(\ref{lab_eqn}) and (\ref{eqnee})
\bea
{\frac{d \sigma^P_{ee}}{d \cos{\theta_L}}} &=& N_P \left\{
(1 + \Delta_{0}^P) \; \left[I^L_0(\theta_L)+I^T_0(\theta_L)\right]
+\Delta_{2,L}^P \; I_2^L(\theta_L)+
\Delta_{2,T}^P \; I_2^T(\theta_L)\right\} \; ,
\label{lab_dist}
\eea
where the functions $I_{0,2}^{\lambda}(\theta_L)$ include the integration 
of the $W$ distributions. In particular,
\bea
I_0^T(\theta_L)&=& 2 \int_{\tau}^{1} \frac{d x}{x}\; 
P_{T}^{W} (x) P_{T}^{W}\left(\frac{\tau}{x}\right)\;
{\cal J}(\theta_L, \beta)
\\
I_0^L(\theta_L)&=& \frac{(r_H-2)^2}{4} 
\int_{\tau}^{1} \frac{d x}{x}\; 
P_{L}^{W} (x) P_{L}^{W}\left(\frac{\tau}{x}\right)\;
{\cal J}(\theta_L, \beta)
\\
I_2^T(\theta_L)&=& 2 \int_{\tau}^{1} \frac{d x}{x}\; 
P_{T}^{W} (x) P_{T}^{W}\left(\frac{\tau}{x}\right)
\;{\cal J}(\theta_L, \beta)\;{\cal F}(\theta_L, \beta)
\\
I_2^L(\theta_L)&=&\frac{(r_H-2)^2}{4} \;\int_{\tau}^{1} \frac{d x}{x}\; 
P_{L}^{W} (x) P_{L}^{W}\left(\frac{\tau}{x}\right)
\;{\cal J}(\theta_L, \beta)\;{\cal F}(\theta_L, \beta) \; ,
\eea
where the factor 2 in the transverse functions $I_{0,2}^T(\theta_L)$
comes from the two different initial $W$ polarization 
transverse projections 
that contribute to a spin-0 state.

For the graviton component, the zeros of the $(1 - 3\cos^2{\theta}) \;$ 
distribution in the $WW$ c.o.m. frame are 
in general shifted by  the  $WW$ boosts to higher values
of $|\cos{\theta}|$. 

\begin{figure}[t]
\vspace*{-5em}
\centerline{\hspace*{3em}
\epsfxsize=9cm\epsfysize=8.0cm
                     \epsfbox{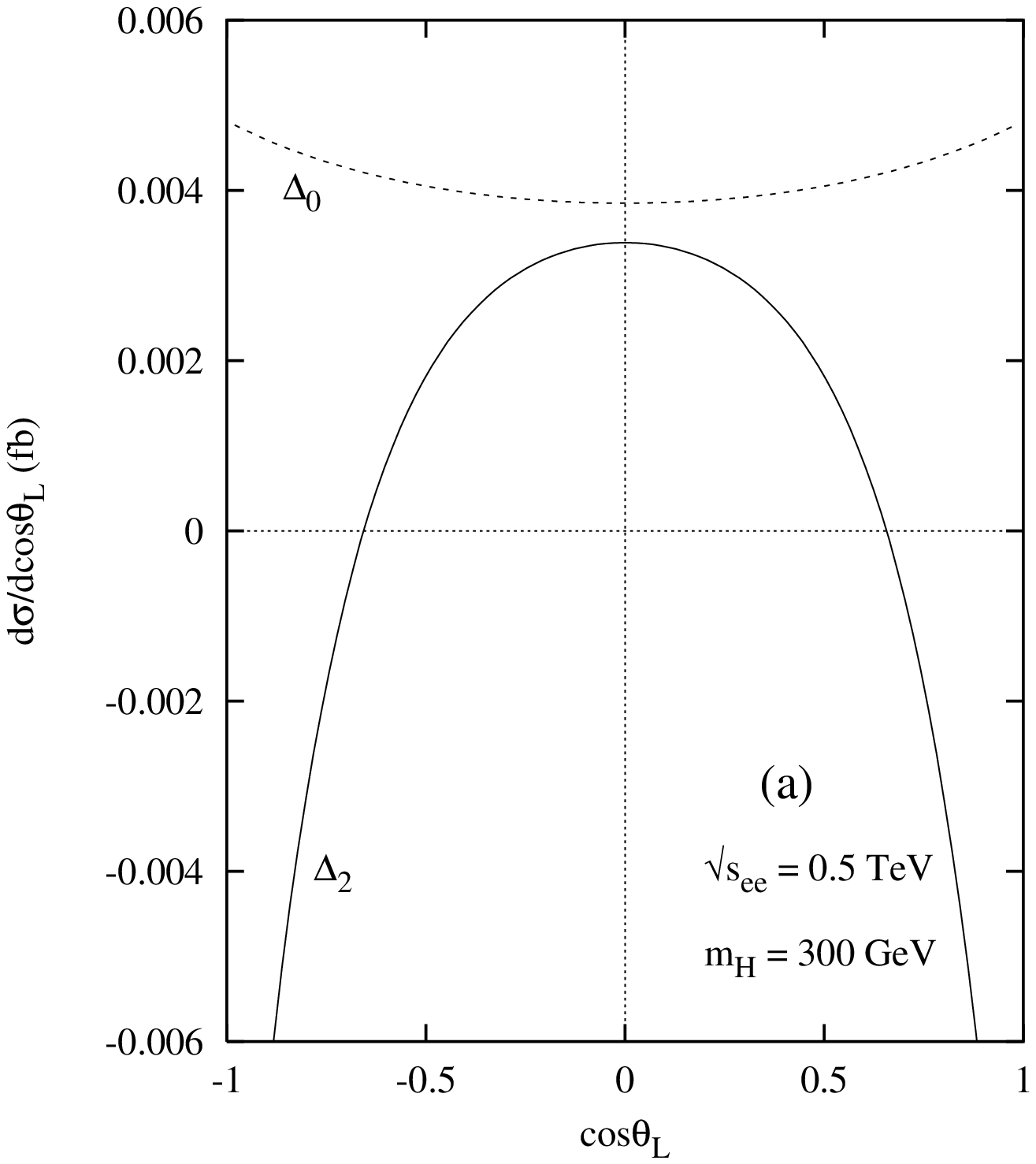}
\epsfxsize=9cm\epsfysize=8.0cm
                     \epsfbox{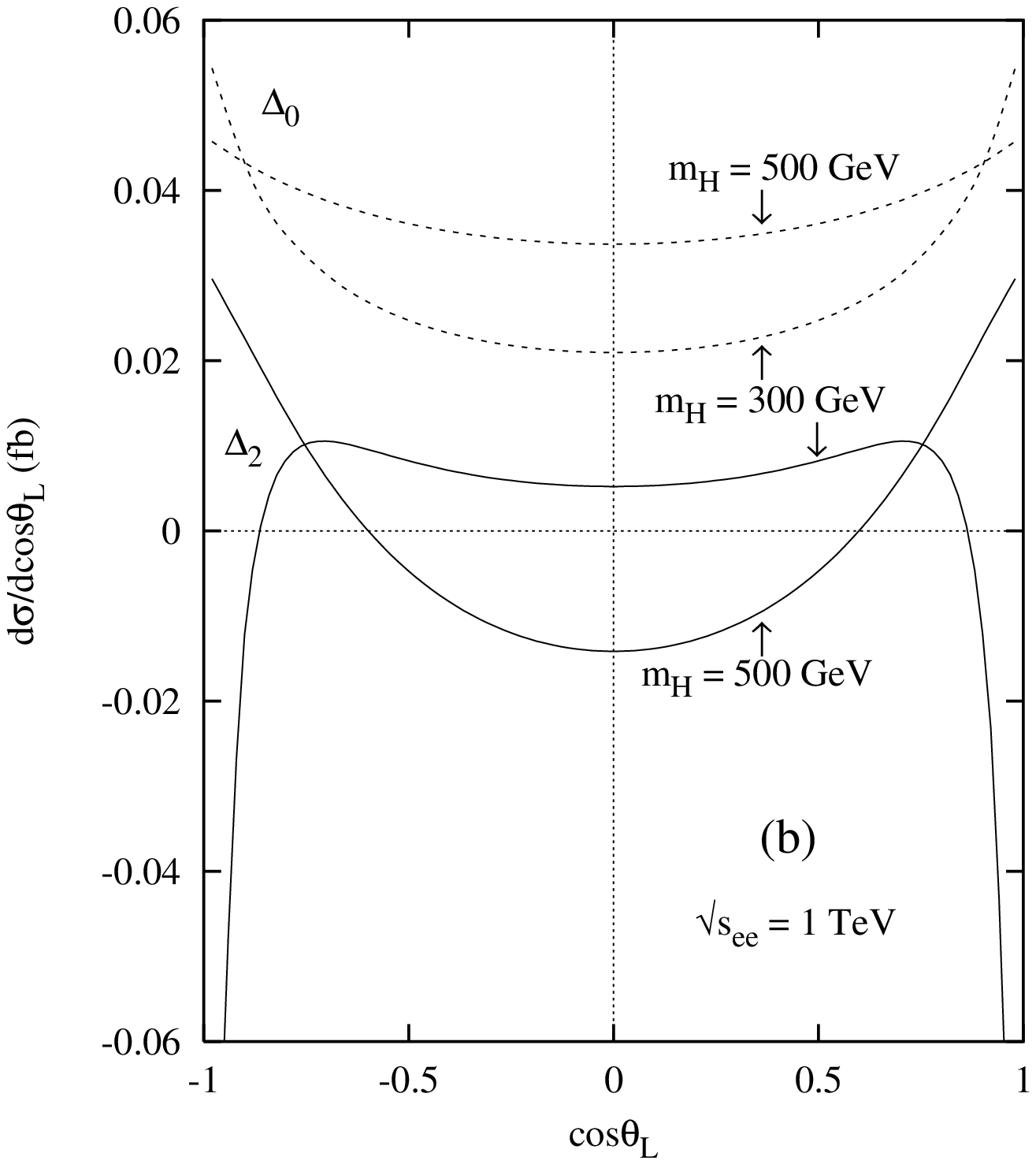}
}
\caption[]{\em 
Graviton (solid line) and
graviscalar (dashed line) contributions to the 
angular distribution in the laboratory frame [Eq.~(\ref{lab_dist})]
for $e^+e^-$ center-of-mass energies 
(a) 500 GeV and (b) 1 TeV.}
\label{distribution}
\end{figure}
In Figure \ref{distribution}a, we show (by the symbol $\Delta_0$) 
the graviscalar contribution 
and (by the symbol $\Delta_2$) the graviton contribution  
(including  both the longitudinal and
 the transverse part) to the total
interference with the SM amplitude in the angular distribution
in Eq.~(\ref{lab_dist}),
at $\sqrt{S}=500$ GeV and $m_H=300$ GeV.

In order to pin down the {\it graviton} contribution 
in Eq.~(\ref{lab_dist}), one can then  measure the angular distribution 
integrated 
over a selected $\cos{\theta_L}$ range, where the corresponding 
distribution
keeps, for instance, a positive (or a negative) value.
The corresponding cross section measured 
with an angular cut  $|\cos{\theta_L}|<\epsilon$ (where $\pm \epsilon$
could be the zeros of the distribution)
will be given by 
\beq
\sigma^{P,\epsilon}_{ee} = \bar{\sigma}^{P,\epsilon}_{ee}\;
\left(1 + \Delta_{0}^P 
+ \alpha^{\epsilon}_L \Delta_{2,L}^{P} + \alpha^{\epsilon}_T \Delta_{2,T}^{P} 
\right) \; ,
\label{crosscut}
\eeq
with
\bea
\bar{\sigma}^{P,\epsilon}_{ee} &=& N_P 
\int_{-\epsilon}^{\epsilon} d\cos{\theta_L}\; 
\left[I^L_0(\theta_L)+I^T_0(\theta_L)\right] \; ,
\nonumber \\
\alpha_{\lambda}^{\epsilon}&=&\frac{\int_{-\epsilon}^{\epsilon} 
d\cos{\theta_L}\;
I^{\lambda}_2(\theta_L)}{
\int_{-\epsilon}^{\epsilon} d\cos{\theta_L}\; 
\left[I^L_0(\theta_L)+I^T_0(\theta_L)\right]} \; ,
\eea
where $\bar{\sigma}^{P,\epsilon}_{ee}$ is the SM cross section restricted 
to the angular range $|\cos{\theta_L}|<\epsilon$ ,
and $\alpha_{\lambda}^{\epsilon}$ are angular  weights for the {\it graviton}
interference effect on the cross section.
Note that, due to Lorentz invariance, one must find
$\alpha_{\lambda}^1=0$. 
Keeping $\epsilon < 1$, on the other hand, will give rise 
in general to nonvanishing
{\it graviton} contributions to the cross sections.

From Figure \ref{distribution}a, by choosing an angular cut equal to the zeros
of the corresponding angular distribution (that is $\epsilon \simeq 0.66$),
 one can optimize the total 
{\it graviton} contribution.
Correspondingly, one  gets the following weights for 
the transverse and longitudinal graviton components : 
$\alpha_{T}\simeq 0.36$
and $\alpha_{L}\simeq 0.16$.

Applying a similar strategy to the $\sqrt{S}=1$ TeV case (cf. Figure 
\ref{distribution}b),
one has, for $m_H=300$ GeV, 
$\epsilon \simeq 0.86$ with $\alpha_{T}\simeq 0.18$
and $\alpha_{L}\simeq 0.09$.
For $m_H=500$ GeV, one can selects the central
range where the graviton distribution
keeps negative values, and set
$\epsilon \simeq 0.60$. Correspondingly, one has $\alpha_{T}\simeq 0.13$
and $\alpha_{L}\simeq 0.43$.

Hence, one in general expects that the graviton coefficients in 
Table~\ref{table1} will contribute to the measured cross section 
with reduction factors of a few tens percent according to 
Eq.~(\ref{crosscut}).
At the same time, the graviscalar contribution, having the same 
flat distribution as the SM signal, will always contribute by the 
total relative amount $\Delta_0$ to both  the total and the cut cross section.

\section{Gravity interference effects at $\mu^+ \mu^-$ colliders}

A  cleaner framework where  to study gravity interference effects 
on the Higgs boson pole is clearly given by a Higgs boson factory.
Although presently challenging from a technological point of view, 
a $\mu^+ \mu^-$ collider with c.o.m. energies around
$m_H$ is the natural place where to realize a Higgs boson factory
\cite{hfactory}.

One then should consider the gravity interference with the 
Higgs exchange diagram for the process
\beq
 \mu^+ \mu^- \rightarrow P \bar P \; ,
\label{mumu}
\eeq
with $P=t,W,Z$.

For  unpolarized initial states,
the cross section for the latter process, including interference 
contributions with the graviscalar and graviton exchange graphs,  can 
be expressed near the Higgs pole as: 
\bea
{\frac{d \sigma^P}{d \cos{\theta}}} &=& 
\frac{\bar{\sigma}^P}{2} \;   
\left\{1 + \Delta_{0}^P + \Delta_{2}^P
 \;(1 - 3\cos^2{\theta}) \right\} \; ,
\label{master_eqn1}
\eea
where
\beq
\bar{\sigma}^{P} ~=~ \frac{d_P}{8 \pi s}\;
\frac{G_F^2 m_{\mu}^2 m_P^4}{(s - m_H^2)^2 + m_H^2 \Gamma _h^2}
\; \sqrt{\frac{s - 4 m_P ^2}{s -4m_{\mu}^2}}\;(s -4m_{\mu}^2)
\; \rho_P\left(\frac{s}{m_P^2}\right) \; ,
\eeq
with 
\beq
\rho_t(x) = (x -4) \;, \;\;\;\;\; 
\rho_{W,Z}(x)=\frac{1}{2} ( x^2 -4x +12 )\; ,
\eeq
is the SM total cross section for the process in Eq.~(\ref{mumu}).
Here, $\sqrt{s}$ is the total collision energy in the $\mu^+ \mu^-$
c.o.m. frame, and the coefficients $d_P$ 
for the different final states have the following values:
$d_t = 3$, $d_W = 1$ and $d_Z = \frac{1}{2}$.

The coefficients $\Delta_{0}^P$ are the same as in Eq.~(\ref{rrr})
\beq
 \Delta_{0}^P = 
 R_\delta\,c_P \left(\frac{\delta -1}{\delta +2}\right) \; ,
\eeq
with $c_t=c_W = \frac{4}{3}$ and $c_Z = \frac{2}{3}$,
and $R_\delta$ as in Eq.~(\ref{rrr}). 
For the coefficients $\Delta_{2}^P$ , one has instead
\beq
\Delta_{2}^t = - \, \frac{4}{3} \; R_\delta \; ,~~~~~~~
\Delta_{2}^{W,Z} =  R_\delta\;\frac{2}{3} \;
\frac{(x-4)(x+6)}{x^2 -4x + 12} \; ,
\eeq
with 
$x = s / m_{W,Z}^2$.

Note that $\Delta_{2}^{t,W}=\Delta_{2,T}^{t,W}$, 
with coefficients $\Delta_{2,T}^{t,W}$ defined as in Eq.~(\ref{rrr}).
As a consequence of the above identities,
a few numerical interesting values of $\Delta_{0}^{t,W}$
and $\Delta_{2}^{t,W}$, for $M_D = 1$ TeV and  $ \delta = 2 $,
can  be found back in the Tables~\ref{table1} and \ref{table2}.

Even in the process $\mu^+ \mu^-\rightarrow P \bar P$
the gravity interference effects can be quite large for high Higgs 
boson masses.
Also in this case, in order to enhance the {\it graviton}
contribution (that vanishes in the total cross section)
it would be sufficient to properly exclude
in the measured cross-section  the forward-backward
direction. This can be straightforwardly done in this case
by properly cutting the $\theta$ range in Eq.~(\ref{mumu}).
  
\section{Conclusions}
In this paper, we computed gravity interference effects
in Higgs boson production at future colliders in the framework of 
the models based on large compact extra dimensions proposed in 
\cite{ADD}.
In particular, we considered the Higgs production
channel via $WW$ fusion at linear colliders 
(that we treat in the   effective $W$ 
approximation) with a subsequent Higgs decay
into pairs of heavy particles ($WW,ZZ,t\bar t$).
We also analyzed Higgs production and decay channels at $\mu^+\mu^-$ Higgs 
factories.
The interference of graviton/graviscalar exchange diagrams
with {\it resonant} Higgs production and decay channels
has the advantage with respect to usual virtual graviton/graviscalar exchange
channels to lead to a completely predictive determination
in terms of the Planck scale $M_D$ and number of extra dimensions
$\delta$. 
The  effect on the SM angular distribution
in general increases with the Higgs boson mass
(for  $\delta=2$, the effect is proportional
to $m_H \Gamma_H$). The {\it graviscalar} interference, that does not
alter the shape of the distributions,  changes its normalization by
a few percent for $m_H > 500$ GeV, if $M_D\simeq 1$ TeV and 
$\delta=2$.

On the other hand, due to the different spin properties of the {\it graviton}
 and Higgs boson amplitude,
the  graviton interference alters the angular shape by
a universal  $(1 - 3\cos^2{\theta})$ distribution 
(in the $W^+W^-$ or $\mu^+\mu^-$ c.o.m. frame) with a coefficient
that is again of the order of a few percent for $m_H > 500$ GeV.
The latter distribution is averaged to zero in the total cross section.
Hence, in order to select a graviton effect, we suggest angular-cut
strategies that enhance the {\it graviton} interference 
contribution in the measured cross section.

In order to detect such indirect graviton effects in Higgs cross section 
measurements, it is  crucial that the actual experimental set up 
will be able to reach the required sensitivity.
While  assessing the final precision 
of  muon colliders is premature at the moment,
 quite a few studies on this subject
have been carried out
for the linear $e^+e^-$ colliders \cite{Aguilar-Saavedra:2001rg}.
In particular, the precision expected on the measurement of the cross section
for Higgs boson production via $WW$ fusion has been considered
in \cite{Desch:2001at} (see also \cite{Dawson:2002wc})
for a light Higgs decaying predominantly into
$b$ quark pairs, and is of the order of a few percent.
A detailed study for heavier Higgs bosons (that are the relevant ones
for our study) is presently missing, to our knowledge.
Anyhow,
a percent precision in the cross section measurements should allow to 
detect some effect at least in the most favorable case
of $M_D\simeq 1$ TeV and 
$\delta=2$ at both linear colliders and Higgs factories.
The effect scales as $\sim 1/M_D^{2+\delta}$ with the Planck mass scale.

A complete treatment (i.e., beyond the effective $W$ approximation)
 of the cross-section in the $WW$ fusion process at linear
colliders is not expected to alter our conclusions.

Note that, by the time experiments at linear colliders
should be operating, the LHC will have 
presumably observed the direct production
of gravitons in the range of parameters that could be 
relevant for our {\it precision}
measurements. In particular, a direct graviton signal is
expected, for $\delta=2,3,4$, for $M_D$ up to a few TeV's \cite{Vacavant:sd,
Vacavant:dd}.
The information derived from the direct graviton production 
and observation at LHC
will definitely help in disentangling the deviations in 
the Higgs cross sections and distributions analyzed in the
present paper.

\section*{Acknowledgments}
We would like to thank M. Giovannini, M. Porrati, and M. Testa
for useful discussions. E.G. and A.D. would also like to thank the Physics
Department of University of Roma ``La Sapienza'', while E.G. thanks also the
CERN Theory Division, for their kind of hospitality during 
the preparation of this work. 
A.D. and E.G. also thank Academy of Finland (project number 48787) 
for financial support.
\vskip 2cm

\end{document}